\documentclass[11pt,a4paper]{article}
\pdfoutput=1
\usepackage{jcappub}

\usepackage{bm}
\usepackage{amsfonts}
\usepackage{subfigure}

\newcommand{\be}{\begin{equation}}
\newcommand{\ee}{\end{equation}}
\newcommand{\bea}{\begin{eqnarray}}
\newcommand{\eea}{\end{eqnarray}}

\begin{document}



\title{Cosmological $N$-body simulations with generic hot dark matter}

\author[a,b]{Jacob Brandbyge,}
\author[a]{Steen Hannestad}

\affiliation[a]{Department of Physics and Astronomy, University of Aarhus, Ny Munkegade 120, DK--8000 Aarhus C, Denmark}
\affiliation[b]{Centre for Star and Planet Formation, Niels Bohr Institute \& Natural History Museum of Denmark, University of Copenhagen, {\O}ster Voldgade 5-7, DK--1350 Copenhagen, Denmark}

\emailAdd{jacobb@phys.au.dk, sth@phys.au.dk}

\abstract{We have calculated the non-linear effects of generic fermionic and bosonic hot dark matter components in cosmological $N$-body simulations. For sub-eV masses, the non-linear power spectrum suppression caused by thermal free-streaming resembles the one seen for massive neutrinos, whereas for masses larger than 1eV, the non-linear relative suppression of power is smaller than in linear theory.  We furthermore find that in the non-linear regime, one can map fermionic to bosonic models by performing a simple transformation.
}

\maketitle

\section{Introduction}

Although many different types of observations constrain the dominant dark matter component of the Universe to be cold, a sub-dominant hot dark matter (HDM) component cannot be excluded. Indeed, standard model neutrinos will inevitably make up a HDM fraction bounded from below by the minimum mass allowed by oscillation experiments. Cosmological structure formation is extremely sensitive to the presence of HDM and therefore cosmology provides by far the strongest current constraint on the mass of standard model neutrinos. Currently cosmological data typically yields a bound on the sum of neutrino masses, $\sum m_\nu$, in the range of 0.1-0.3 eV depending on which data sets, parameters, and priors are used (see e.g.\ 
\cite{Ade:2015xua,Moresco:2016nqq,Vagnozzi:2017ovm,Giusarma:2016phn,Couchot:2017pvz,Capozzi:2017ipn}).

The effect of massive standard model neutrinos on non-linear structure formation has been studied extensively in the literature. Most notably massive neutrinos have been incorporated into $N$-body simulations using a variety of different approaches
(see e.g.\ \cite{Brandbyge:2008rv,Brandbyge:2009ce,Brandbyge:2010ge,Brandbyge:2008js,Massara:2014kba,Agarwal:2010mt,Viel:2010bn,Wagner:2012sw,AliHaimoud:2012vj,Baldi:2013iza,Inman:2015pfa,Lawrence:2017ost,Bird:2011rb}), and good accuracy has been reached on observables such as the total matter power spectrum over a wide range of scales and neutrino masses.

However, many extensions of the standard model also predict the presence of light, weakly interacting particles which can contribute to the dark matter. Perhaps the most obvious example is the light sterile neutrino. If its mass is in the keV range it can even be the dominant dark matter component, but for masses in the eV range it is constrained to be sub-dominant. Other possibilities for light dark matter is eV axions or majorons.

Using linear perturbation theory HDM is mainly constrainable via its effect on very large scales, i.e.\ the Cosmic Microwave Background (CMB) or Baryon Acoustic Oscillations (BAO). In this case HDM mainly affects structure formation via two parameters: Its current contribution to the cosmic energy density, $\Omega_X h^2$, and its contribution to the relativistic energy density at early times, $N_{\rm eff}$. However, even though these two parameters are measurable the mass of the HDM particle itself is harder to constrain.

In non-linear structure formation, however, the physical HDM particle mass becomes crucially important. Models with almost identical large scale behaviour become very different in the non-linear regime and this in turn provides a possible means of distinguishing different types of HDM. In order to have the same $\Omega_X h^2$, $N_{\rm eff}$ must be decreased when the physical particle mass increases. This means that the distribution becomes colder when the mass is increased thus facilitating much more efficient infall into existing cold dark matter (CDM) potential wells. This effect can be quite dramatic for even quite moderate physical particle masses.

In this work we study structure formation in the non-linear regime for various different HDM particle masses and clearly demonstrate that differences in the physical particle mass leads to very different clustering behaviour in the non-linear regime.

The paper is structured as follows: In section \ref{sec:linear_theory} we discuss HDM in linear perturbation theory as well as current observational constraints. In section \ref{sec:methods} we describe the numerical setup we use to study HDM clustering in the non-linear regime, and in section \ref{sec:results} we present our main results. Finally, section \ref{sec:conclusions} contains a discussion and summary of the main results.

\section{Hot dark matter in linear theory}\label{sec:linear_theory}

\subsection{Power spectrum suppression}
\begin{figure}[t]
   \vspace*{-2.8cm}
\begin{center}
\hspace*{-0.5cm}
\includegraphics[width=14cm]{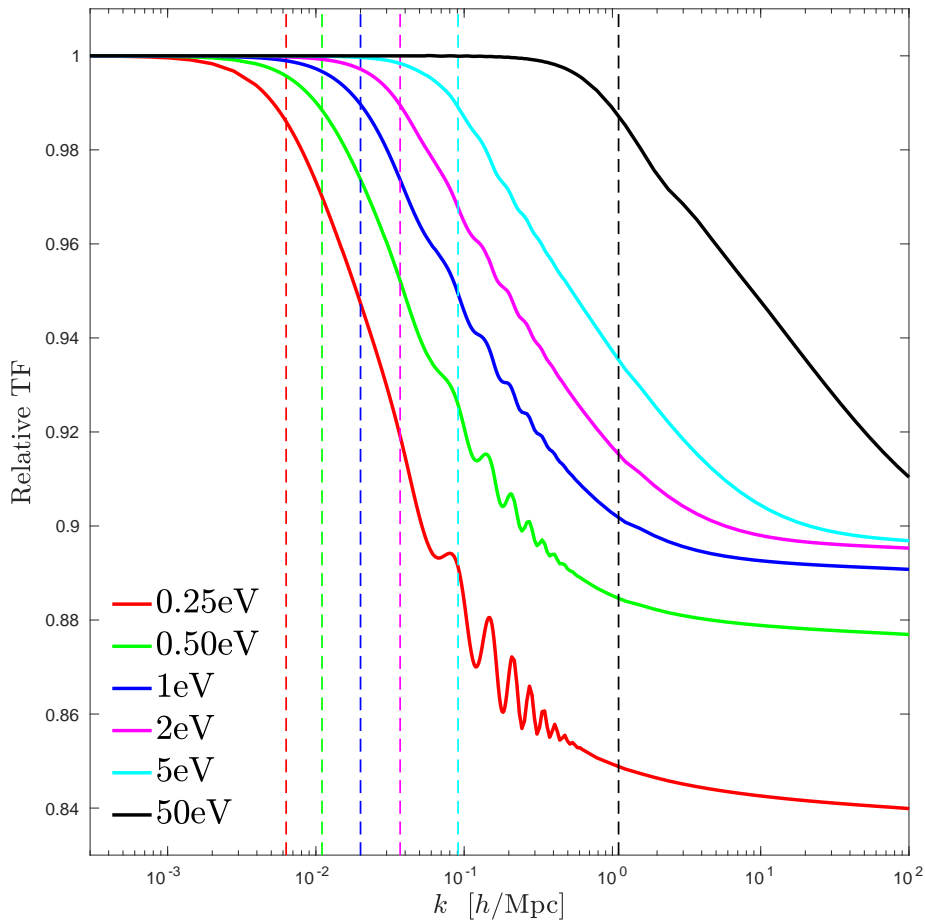}
\end{center}
 \vspace*{-3.0cm}
\caption{The figure displays the relative transfer function at $z=0$ between models with various hot dark matter masses and a model with cold dark matter only. The constant suppression seen for each mass at small scales is caused by different values of $N_{\rm eff}$. A larger value of $N_{\rm eff}$ leads to a longer radiation dominated period, which in turn suppresses the power further for all scales inside the horizon at matter-radiation equality. The vertical dashed lines show the free-streaming scale for each model, as explained in the main text.}
   \label{fig:linear}
\end{figure}

The effect of HDM in the form of neutrinos has been studied numerous times in the literature. In essence, the presence of HDM causes suppression of the matter power spectrum on scales below the free-streaming scale. In Fig.~\ref{fig:linear} we show a number of HDM models. These are all characterised by having $\Omega_X = 0.005$, but different contributions to $\Delta N_{\rm eff}$. $\Delta N_{\rm eff}$ is defined as
\begin{equation}
\Delta N_{\rm eff} \equiv \left. \frac{\rho_X}{\frac{7}{8} \left(\frac{4}{11}\right)^{4/3} \rho_\gamma} \right|_{T \gg m_X},
\end{equation}
where $\rho_\gamma$ is the density in photons, and we have assumed the HDM to be a thermalised fermion, but with a temperature, $T$, different from that of standard model neutrinos. The relation between $\Omega_X h^2$, $\Delta N_{\rm eff}$, and $m_X$ in this case becomes
\begin{equation}
\Omega_X h^2 \simeq (\Delta N_{\rm eff})^{3/4} \frac{m_X}{94.1 \, {\rm eV}}.
\end{equation}
The models from left to right in Fig.~\ref{fig:linear} have $\Delta N_{\rm eff} = 0.90$, $0.36$, $0.14$, $0.056$, $0.017$, $7.7 \times 10^{-4}$, so that the physical particle masses correspond to 0.25, 0.5, 1, 2, 5, and 50eV, respectively.
To a reasonable approximation the power suppression on scales much smaller than the free-streaming scale can for standard model neutrinos be approximated as $\Delta P/P \sim -8 \Omega_X/\Omega_m$ \cite{Hu:1997mj,Lesgourgues:2006nd,Hannestad:2006zg,Wong:2011ip}.
The model presented here is somewhat different because it assumes 3 very light standard model neutrinos, and an additional HDM component.
This difference becomes apparent when we compare the models with $m_X = 0.25$ and 0.5 eV to the others. In these cases the total effective number of relativistic degrees of freedom until after matter-radiation equality is significantly higher than the standard model 3.046. This delays matter-radiation equality and leads to an additional suppression of power on all scales inside the horizon at matter-radiation equality (as well as a shifting of the $k$-value corresponding to the horizon size).

Fig.~\ref{fig:linear} also illustrates the effect of the free-streaming scale.
We define the free-streaming length scale for a non-interacting species as
\begin{equation}
d_{\rm FS}(a=1) = \int_0^1 \frac{\langle v \rangle da}{a^2 H},
\end{equation}
and the corresponding wavenumber as $k_{\rm FS} = \frac{2 \pi}{d_{\rm FS}}$.
Here we have used the mean velocity of the distribution, defined as $\langle v \rangle = \int (p/E) f(p) d^3 p/\int f(p) d^3p$.
For the cases in the figure the free-streaming wavenumbers are
$k_{\rm FS} = 6.3 \times 10^{-3}$, 0.011, 0.020, 0.037, 0.091, and 1.1$h/{\rm Mpc}$, respectively. This can be seen in the figure to be approximately the location where power suppression sets in.
We note here that our definition of the free-streaming scale is slightly different from what is used in e.g.\ Lesgourgues and Pastor
\cite{Lesgourgues:2006nd}.

\subsection{Phase space distribution of the hot dark matter}\label{sec:fermionvsboson}

Our prototype HDM component is a fully thermalised fermion. However, the HDM could consist of bosons, or could have a non-thermal distribution.
For most HDM candidates we would expect linear theory observables to be sensitive to at most a few effective parameters describing their distributions. 
As an example, consider a fully thermalised bosonic species with a distribution function parameterized by $g_{s,b}$, $T_{{\rm eff},b}$, and $m_b$. The energy densities in the relativistic and non-relativistic limits are given by
\begin{eqnarray}
\rho_{{\rm R},b} & = & \frac{\pi^2}{30} g_{s,b} T_{\rm eff,b}^4, \\
\rho_{{\rm NR},b} & = & \pi^2 \zeta(3) g_{s,b} m_b T_{\rm eff,b}^3,
\end{eqnarray}
respectively. This can be mapped into a fermionic dark matter candidate of equal mass with the same relativistic and non-relativistic energy densities given by:

\begin{eqnarray}
\rho_{{\rm R},f} & = & \frac{7}{8} \frac{\pi^2}{30} g_{s,f} T_{\rm eff,f}^4, \\
\rho_{{\rm NR},f} & = & \frac{3}{4} \pi^2 \zeta(3) g_{s,f} m_{f} T_{\rm eff,f}^3, 
\end{eqnarray}
where $m_f = m_b$, $g_{s,f} = 343 g_{s,b}/162$, $T_{{\rm eff},f} = 6 T_{{\rm eff},b}/7$ follows from requiring that $\rho_{{\rm R},f} = \rho_{{\rm R},b}$ and $\rho_{{\rm NR},f} = \rho_{{\rm NR},b}$, i.e.\ requiring that the energy densities match in both the relativistic and non-relativistic limits. This also has the feature of giving almost exactly the same redshift for the transition from the relativistic to the non-relativistic regime.

Although this would not correspond directly to a physical fermion state, because $g_{s,f}$ is unphysical, it illustrates the point that linear theory observables are sensitive only to a few effective parameters describing the asymptotic energy densities and the epoch of the transition from the relativistic to the non-relativistic regime (see e.g.\ \cite{Hannestad:2005bt} for a more detailed discussion of the linear theory differences).

The conclusion is that a thermally distributed boson species can be mapped to a fermion species using only three effective parameters, and that this effective fermion model provides linear theory observables which are {\it de facto} indistinguishable from those of the original boson model.

However, even though this is true in linear theory it is plausible that this could be different for observables probing non-linear structure formation. Indeed, one would expect the low energy tails of the particle velocity distributions to be very different for fermions and bosons. This in turn leads to very different clustering in the central parts of halos (see e.g.\ \cite{Hannestad:2005bt} and references therein). 
In section \ref{sec:results_fvb} we will study possible differences between the fermionic and bosonic cases in the non-linear regime.

\subsection{Current constraints from linear theory}
HDM is severely constrained to make up at most a small fraction of the total dark matter density by current structure formation data.
In order to get an idea about which models are interesting to study with detailed $N$-body simulations we have performed a standard parameter estimation analysis on the standard $\Lambda$CDM model with 3 massless neutrinos plus an additional dark matter component (implemented as a sterile neutrino, i.e.\ a thermalised fermion) parameterised using the physical particle mass, $m_X$, and the contribution to the relativistic energy density at early times, $\Delta N_{\rm eff}$. 
We note that this is slightly different from the parameterisation normally used in which the two parameters are $m_{\rm eff,X}$ and $\Delta N_{\rm eff}$. We also use flat priors on both $\log_{10}m_X$ and $\log_{10}(\Delta N_{\rm eff})$ which has the effect of shifting the preferred values down.

In terms of data we use the Planck 2015 data, including high-$l$ $E$-polarisation \cite{Ade:2015xua,Aghanim:2015xee} (the same combination as in \cite{Archidiacono:2016kkh}).  We also include BAO data from a variety of
different surveys: 6dFGS~\cite{Beutler:2011hx}, SDSS-MGS~\cite{Ross:2014qpa}, BOSS-LOWZ \cite{Anderson:2012sa} and
CMASS-DR11~\cite{Anderson:2013zyy}. To perform parameter estimation and derive constraints we have used
the publicly available CosmoMC code \cite{Lewis:2002ah}.

Using our parameterisation we find no bound on $m_X$ at 95\% C.L., in accordance with expectations. At 68\% we formally find a bound on $m_X$ of $\log_{10}(m_X) < -0.23$. However, this bound is entirely caused by prior volume, i.e.\ the fact that the allowed range in $\Delta N_{\rm eff}$ shrinks when $m_X$ increases.
The formal bound on $\Delta N_{\rm eff}$ is $\log_{10}(\Delta N_{\rm eff}) < -0.64$ ($\Delta N_{\rm eff} < 0.23$) at 95\%. The constraint found in \cite{Archidiacono:2016kkh} using approximately the same data is somewhat less restrictive ($\Delta N_{\rm eff} < 0.4$), and the difference is again caused by our use of logarithmic priors.

 \begin{figure}[t]
\begin{center}
\includegraphics[width=14cm]{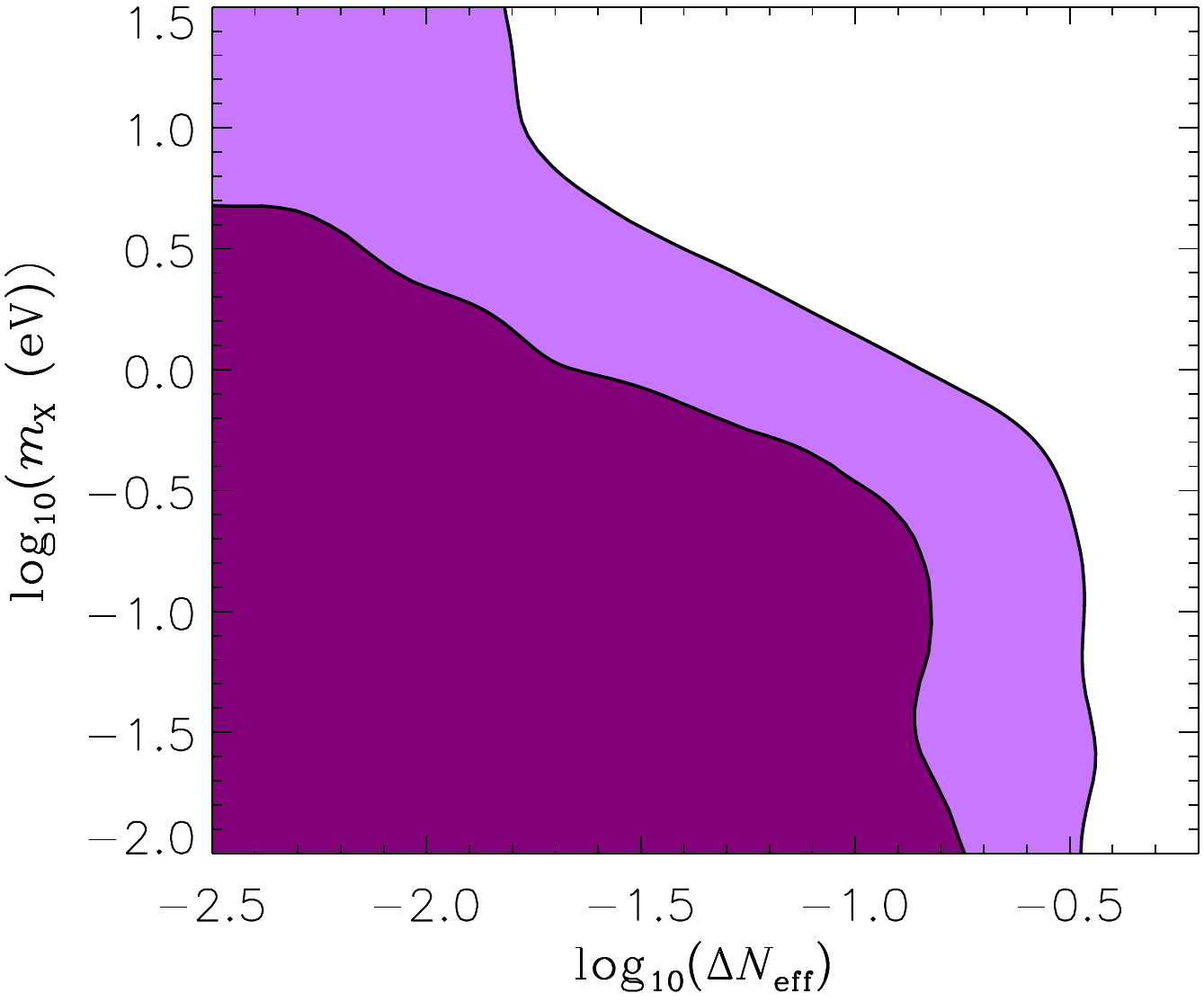}
\end{center}
\caption{Current constraints on $\Delta N_{\rm eff}$ and $m_X$ from CMB and BAO data.}
   \label{fig:cosmomc}
\end{figure}

\begin{table}[t]
    \begin{center} 
        \begin{tabular}{c c c c c c c c} 
           \hline
           Sim &  $m_X$(eV) & $\Delta N_{\rm eff}$ & $T_{X,0}$(K)  & $N_{\rm part}^c$ &  $N_{\rm part}^{\rm X}$ &Type&Distribution \\           
          \hline
          A&      0             & 0                                  & 0               & $512^3$                                & 0                   & Fermion & FD \\  
          B&      0.25        &  0.8976                        & 1.8935       & $512^3$                                 & $512^3$   & Fermion & FD \\
          C&      0.50        &  0.3562                         &  1.5029        & $512^3$                                 & $512^3$ & Fermion & FD   \\
          D&      1             &  0.1414                        &  1.1929        & $512^3$                                 & $512^3$  & Fermion & FD  \\
          E&       2            &  0.05610                      &  0.9468        & $512^3$                                 & $512^3$  & Fermion & FD  \\
          F&       5             &  0.01653                     &  0.6976        & $512^3$                                 & $512^3$  & Fermion & FD  \\
          G  &       50           &  $7.674\cdot 10^{-4}$    &  0.3238         &$512^3$                                 & $512^3$   & Fermion & FD  \\          
          AA &      0             & 0                                & 0                  & $1024^3$                                & 0               & Fermion & FD  \\  
          BB  &      0.25        &  0.8976                       & 1.8935         & $1024^3$                               & $1024^3$& Fermion & FD    \\
          CC  &      0.50       &  0.3562                         &  1.5029        & $1024^3$                               & $1024^3$& Fermion & FD    \\
          DD  &      1          &  0.1414                         &  1.1929        & $1024^3$                                & $1024^3$& Fermion & FD    \\
          EE  &       2          &  0.05610                       &  0.9468        & $1024^3$                                & $1024^3$& Fermion & FD    \\
          FF &       5          &  0.01653                       &  0.6976        & $1024^3$                                & $1024^3$ & Fermion & FD   \\
          GG &       50         &  $7.674\cdot 10^{-4}$        &  0.3238        & $1024^3$                                & $1024^3$   & Fermion & FD  \\           
          R&       2            &  0.05610                      &  0.9468        & $512^3$                                 & $512^3$   & Fermion & FD \\
          S&       2            &  0.05610                      &  1.1046        & $512^3$                                 & $512^3$ & Boson & BE   \\
          T&       2            &  0.05610                      &  0.9468        & $512^3$                                 & $512^3$ & Fermion & BE   \\                                     \hline
      \end{tabular}
      \end{center}
          \caption{The table shows parameters and initialization methods for the $N$-body simulations used in this work.  $m_X$ denotes the HDM particle mass, $\Delta N_{\rm eff} \equiv N_{\rm eff}-3.046$, $T_{X,0}$(K) is the HDM temperature today, and $N_{\rm part}^c$ and $N_{\rm part}^X$ are the numbers of CDM and HDM $N$-body particles, respectively. $Type$ denotes the particle type assumed in the linear theory evolution, and $Distribution$ labels the thermal velocity distribution used in the $N$-body simulation. All transfer functions are calculated with CAMB except for the simlations R, S, and T, where we have used CLASS. All simulations have a box size of $512 {\rm Mpc} / h$ and an $N$-body starting redshift of 49.} 
      \label{table:nbody_sims} 
\end{table}
\section{Simulation methods} \label{sec:methods}
\subsection{Cosmology and initial conditions}
All our simulations are presented in Table~\ref{table:nbody_sims}. The linear theory initial conditions are calculated with CAMB \cite{Lewis:1999bs} except for simulations R, S, and T, where we used CLASS \cite{Lesgourgues:2011re,Blas:2011rf,Lesgourgues:2011rg,Lesgourgues:2011rh}. The following cosmology has been assumed: $\Omega_m =  0.3$, $\Omega_b= 0.05$, $\Omega_c = \Omega_m - \Omega_b-\Omega_X$ for the matter, baryonic and CDM density parameters, respectively, and where $\Omega_X$ is the HDM density parameter. We furthermore assume three massless neutrinos, a cosmological constant with $\Omega_\Lambda = 0.7$, a scalar spectral index of $n_s=1$ and a normalisation given by $A_s = 2.3\cdot 10^{-9}$.

Our HDM simulations have an effective mass fixed by $m_{\rm eff} = 94.1  \Omega_X h^2$eV. Since we use $\Omega_X = 0.005$, we get $m_{\rm eff}=0.23{\rm eV}$, which is close to the current upper bound inferred from various cosmological datasets \cite{Ade:2015xua}. For a given HDM mass ($m_X$) $\Delta N_{\rm eff}$ and $T_X$, the HDM temperature, are fixed by the relations $\Delta N_{\rm eff} = \left(\frac{m_{\rm eff}}{m_X}\right)^{4/3}$ and $T_X = T_\nu  (\Delta N_{\rm eff})^{1/4}$, where $T_\nu$ is the standard model neutrino temperature.

From $z=49$ onwards we solve the non-linear evolution of the density perturbations with \textsc{gadget}-2 \cite{Springel:2005mi}. Since we simulate a multi-component fluid, with HDM masses as high as $50$eV, we initialize the $N$-body particle positions and velocities with the Zel'dovich Approximation \cite{Zeldovich:1969sb} only, and neglect higher order corrections \cite{Crocce:2006ve}. This is, though, more than adequate for our purposes. Again, since we have a multi-component fluid, the initial $N$-body particle velocities are found from the difference of two $N$-body particle position grids, centered at $z=49$.

In the $N$-body simulation the baryons are treated as CDM, but the initial conditions are generated from a weighted sum of baryon and CDM transfer functions. Radiation perturbations (massless neutrinos and photons) are neglected, since, as was recently shown in \cite{Brandbyge:2016raj, Adamek:2017grt}, they do not affect the non-linear regime, when the $N$-body simulations are initialized after $z=99$.

\subsection{Different hot dark matter simulation methods}
In all the simulations the HDM perturbations are discretized with $N$-body particles. In addition to the bulk velocity, these particles receive a thermal velocity drawn from relativistic Fermi-Dirac (FD) or Bose-Einstein (BE) distributions. This method was used to simulate sub-eV neutrinos in \cite{Brandbyge:2008rv}, see also \cite{Viel:2010bn, Agarwal:2010mt,Bird:2011rb}.

Other methods to simulate light massive neutrinos / HDM include: A grid-based approach, where the neutrinos are represented as a fluid in Fourier space and evolved with linear theory only \cite{Brandbyge:2008js}, or with a feedback from the non-linear gravitational field to the linear neutrino equations \cite{AliHaimoud:2012vj}. Finally, a hybrid method exists, where neutrinos initially reside on a linear Fourier grid from where they are later converted to particles. This hybrid approach was explored in \cite{Brandbyge:2009ce}, and subsequently used to calculate halo properties for a massive neutrino cosmology in \cite{Brandbyge:2010ge}.

The HDM mass range covered in this paper, could point towards the use of the hybrid approach for the lowest masses, and the particle method for the higher masses. For consistency though, we have chosen the particle method exclusively.

\subsection{The correlation of bulk and thermal velocities}
At the $N$-body starting redshift of $z=49$ a thermal velocity is added to the $N$-body particle bulk velocity, under the assumption that they are uncorrelated. This is justified when the bulk component is sub-dominant, so that density perturbations cannot significantly have altered the total (bulk + thermal) velocity component away from an FD distribution. This is the case for sub-eV neutrinos.

But for the higher mass neutrinos this assumption breaks down. For $50$eV masses and $z=49$ the bulk component in a $512 {\rm Mpc} / h$ box (the bulk velocity is significantly dominated by large-scale modes) is roughly a factor of 5 larger than the thermal component (which is at the $20$km$/s$ level). Dynamically, the lower velocity part of the total velocity distribution which was primordially Fermi-Dirac is no longer so.

In a worst case scenario, all the low thermal velocity HDM particles have moved into the overdensities. But since the CDM perturbations are at the $1\%$ level at $z=49$ we are far from this scenario. This means that the HDM particles still locally have a primordial thermal component which is FD distributed, with roughly the same temperature everywhere. These considerations do not hold, at late times, in the very non-linear regime. 

So, as long as the perturbations are linear, it is justified to assume the same Fermi-Dirac distribution everywhere; regardless of the ratio between typical bulk and thermal velocities. In the non-linear regime, this assumption is only justified for very low mass HDM particles, which are homogeneously distributed. Higher mass HDM particles will segregate, with the ones with a low thermal velocity falling into structures, and the remaining ones being more evenly distributed. In this case the primordial thermal component is locally not Fermi-Dirac. In sum, the error incurred goes like $\delta$ and not $v_{\rm bulk} / v_{\rm thermal}$.

 \begin{figure}[t]
  \vspace*{-3.8cm}
\begin{center}
\hspace*{-1.6cm}
\includegraphics[width=18cm]{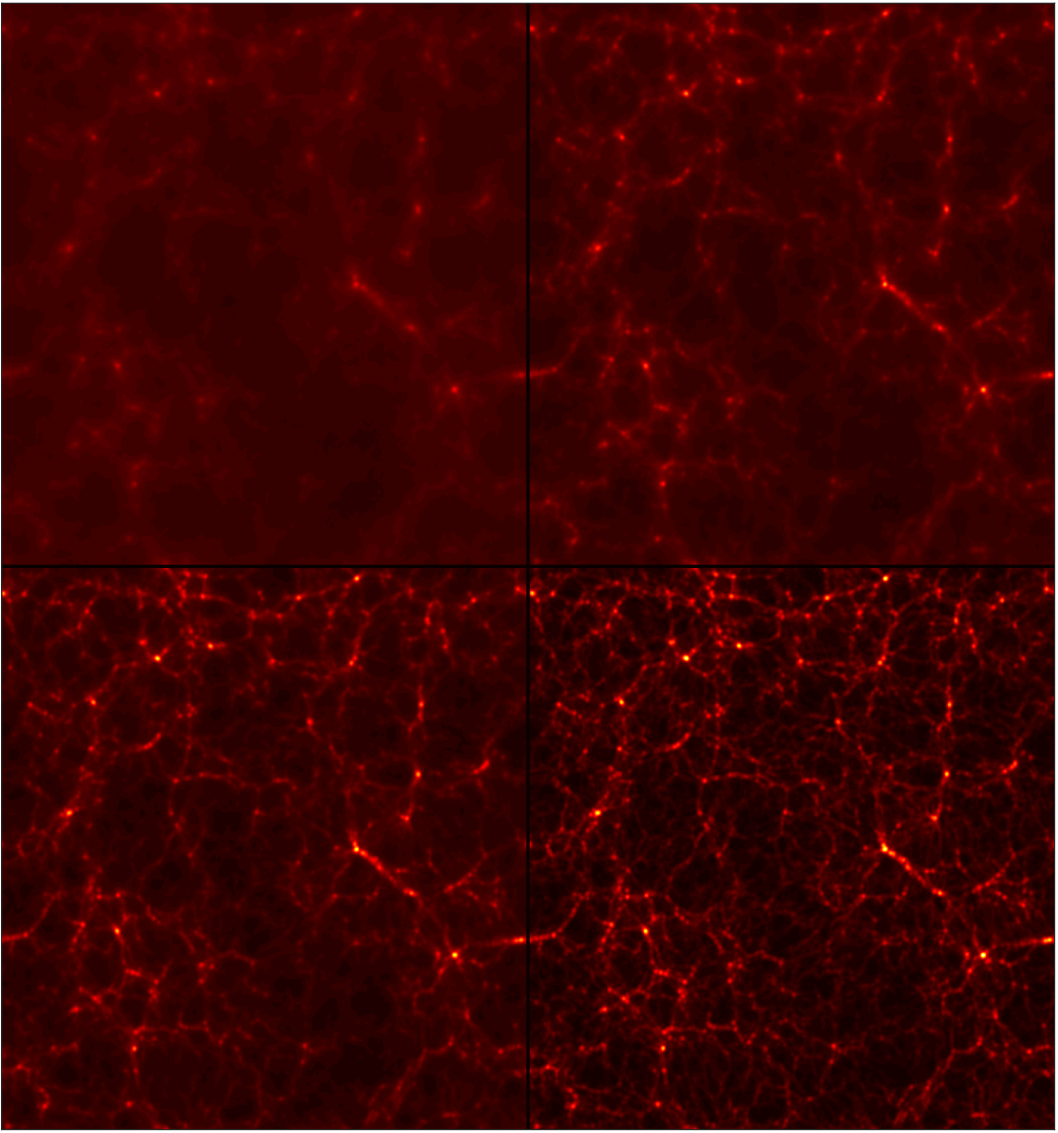}
\end{center}
 \vspace*{-4cm}
\caption{The figure displays $20 {\rm Mpc} /h$ thick slices of the $512 {\rm Mpc} / h$ cubed simulation volume (scaled as $\sqrt{\delta+1}$) at $z=0$. Top left panel shows the $0.25 {\rm eV}$ HDM distribution (simulation BB), top right the $0.5 {\rm eV}$ HDM distribution (simulation CC), bottom left the $1 {\rm eV}$ HDM distribution (simulation DD), and finally the bottom right panel shows the CDM distribution from a CDM only simulation (AA). The images were produced by using the adaptive smoothing length kernel of \cite{Monaghan:1985}.}
   \label{fig:rho_all}
\end{figure}

 \begin{figure}[t]
  \vspace*{-2.5cm}
\begin{center}
\hspace*{-1.0cm}
\includegraphics[width=18cm]{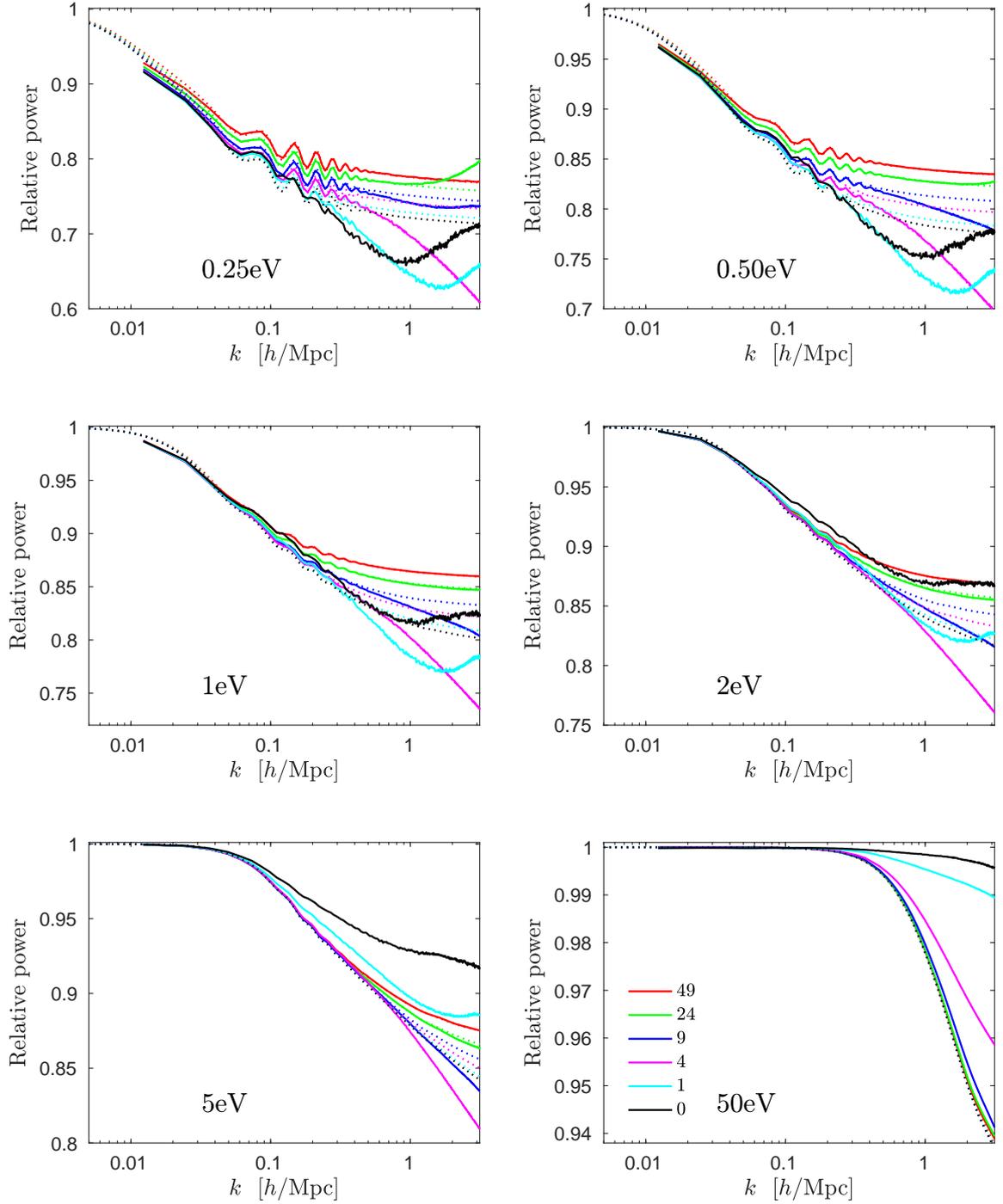}
\end{center}
 \vspace*{-2.5cm}
\caption{The figure displays how hot dark matter affect the relative matter power spectrum when compared to a model with cold dark matter only. Each image shows the suppression at 6 different redshifts. Linear theory is represented by dotted lines and non-linear theory by solid lines. All statistics is taken from the $1024^3$ particle simulations, see Table~\ref{table:nbody_sims}.}
   \label{fig:power_suppression}
\end{figure}

 \begin{figure}[t]
  \vspace*{-3.0cm}
\begin{center}
\hspace*{-0.6cm}
\includegraphics[width=12cm]{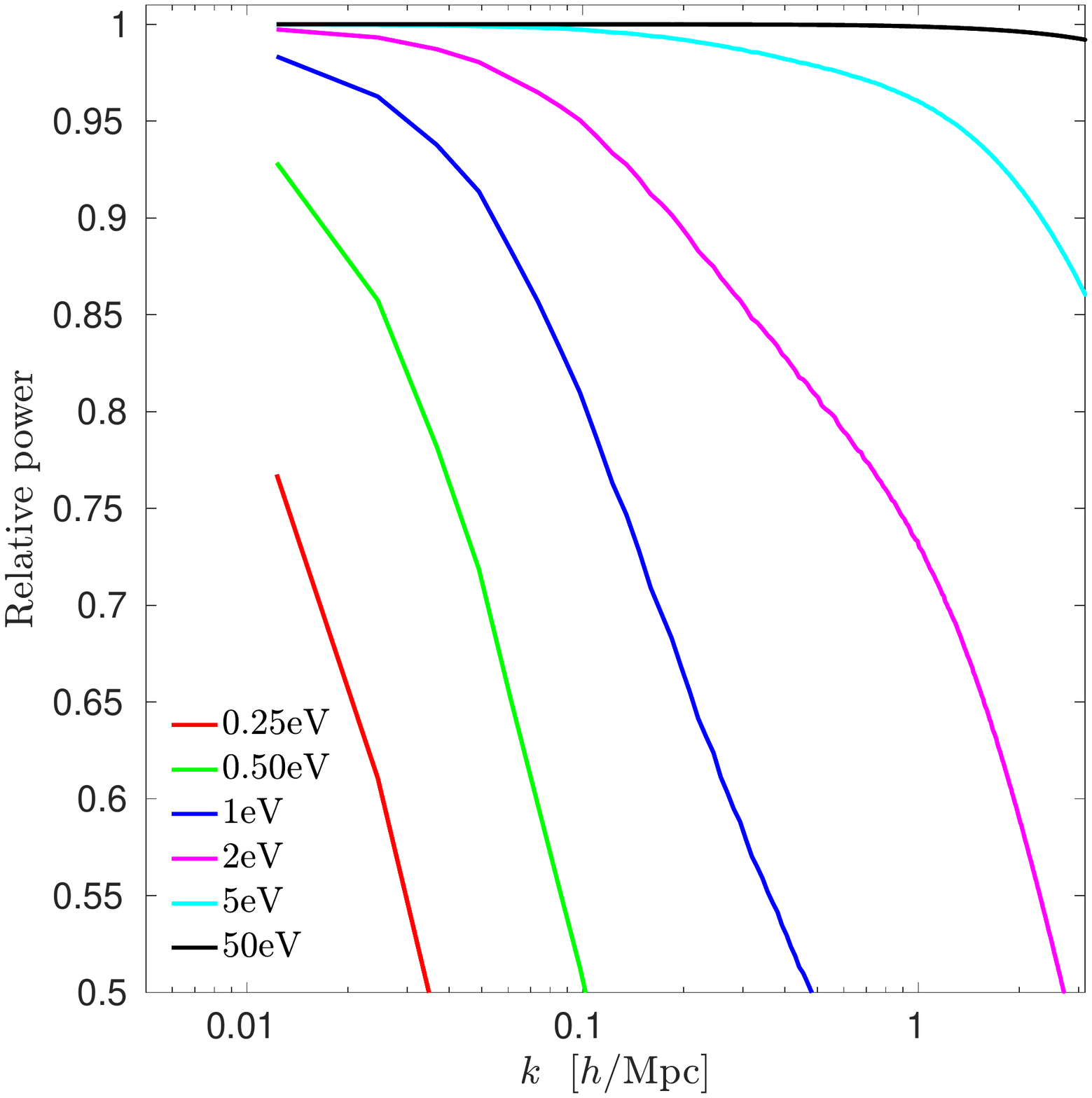}
\end{center}
 \vspace*{-2.6cm}
\caption{The figure displays HDM power spectra divided by the CDM power spectrum from the same simulation at $z=0$. All statistics is taken from the $1024^3$ particle simulations.}
   \label{fig:power_nu_div_cdm}
\end{figure}
\section{Non-linear results}\label{sec:results}
Fig.~\ref{fig:rho_all} shows $20 {\rm Mpc} / h$ thick slices of the $512 {\rm Mpc} / h$ cubed simulation volume. The upper two panels display $0.25{\rm eV}$ and $0.5 {\rm eV}$ HDM distributions, and the lower panels show the results for $1{\rm eV}$ HDM and the CDM distribution from a pure CDM simulation. The effect of thermal free-streaming of the HDM particles is clearly visible, but for masses larger than $\sim 1{\rm eV}$, the density distributions become visually identical to the CDM distribution.  

\subsection{Generic hot dark matter power spectrum suppression}
Fig.~\ref{fig:power_suppression} shows the relative decrease in matter power as a function of redshift for a range of HDM particle masses. In all cases the power suppression is measured relative to a pure CDM simulation. Both linear (dotted lines) and non-linear (solid lines) theory is displayed.

The redshift evolution of the suppression depends significantly on the HDM particle mass. For the sub-eV particle masses the evolution pattern is reminiscent of the evolution for a standard massive neutrino cosmology (see \cite{Brandbyge:2008rv}): Due to the larger amount of clustering in the pure CDM simulation, perturbations of a given scale collapse earlier than they do in a cosmology where part of the CDM component is replaced with a HDM component. Eventually the scale in the mixed CDM-HDM simulation collapses which produces the turnover in the relative power spectrum.

Since the value of $N_{\rm eff}$ differs from the one with 3 massive neutrinos, the maximum suppression today cannot be fitted by the formulas $1-8~\Omega_X / \Omega_m$ in the linear regime and $1-10~\Omega_X / \Omega_m$ in the non-linear regime \cite{Brandbyge:2008rv}.

The turnover in the relative power spectrum for the $0.25 {\rm eV}$ simulation at $z=24$ is a noise term related to the finite number of HDM $N$-body particles. This feature is not related to the non-linear evolution.

For the higher HDM particle masses the redshift evolution of the relative power spectrum is markedly different. For the $2{\rm eV}$ and $5{\rm eV}$ cases, the evolution initially resembles the one for the lower mass simulations, but at later times non-linear theory predicts a significantly smaller suppression in the relative power spectrum than do linear theory. In the $50{\rm eV}$ case, this latter evolution sets in much earlier.

This smaller non-linear suppression can be understood from the following considerations: The relative power spectrum for, say, the $50{\rm eV}$ mass is basically unity for $k\lesssim 0.3 h/{\rm Mpc}$. Since power in the non-linear regime is moved from larger to smaller scales (see \cite{Brandbyge:2017wyw}), this means that in the simulations with / without HDM a roughly equal amount of power is moved to smaller scales (from the modes with $k<0.3 h/{\rm Mpc}$). Therefore, as time evolves the relative power spectrum approaches unity at progressively smaller scales. Eventually, the relative power spectrum is unity for $k\lesssim 1 h/{\rm Mpc}$ at $z=0$ for the $50{\rm eV}$ case.

Fig.~\ref{fig:power_nu_div_cdm} shows, for a given simulation, the ratio of the HDM power spectrum to the CDM power spectrum at $z=0$. The dependence of thermal free-streaming on the HDM particle mass is clearly visible. 

 \begin{figure}[t]
  \vspace*{-7.5cm}
\begin{center}
\hspace*{-1.0cm}
\includegraphics[width=18cm]{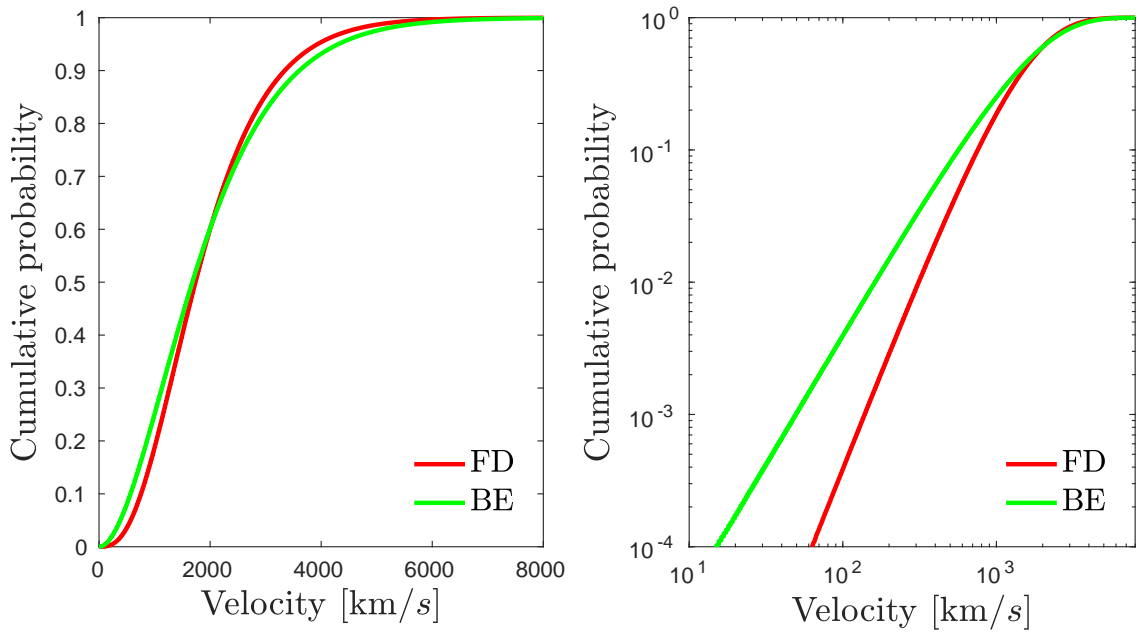}
\end{center}
 \vspace*{-7.0cm}
\caption{The panels show the cumulative velocity distribution functions at $z=49$ for $2{\rm eV}$ fermions and bosons with the temperature relation $T_{{\rm eff},b} = \frac{7}{6} T_{{\rm eff},f}$. Both panels show the same distribution functions, but the figure on the right has logarithmic axis, which more clearly demonstrates the higher weight of the Bose-Einstein distribution at the low velocity end.}
   \label{fig:velocity_dist2}
\end{figure}

 \begin{figure}[t]
  \vspace*{-4.0cm}
\begin{center}
\hspace*{-1.0cm}
\includegraphics[width=18cm]{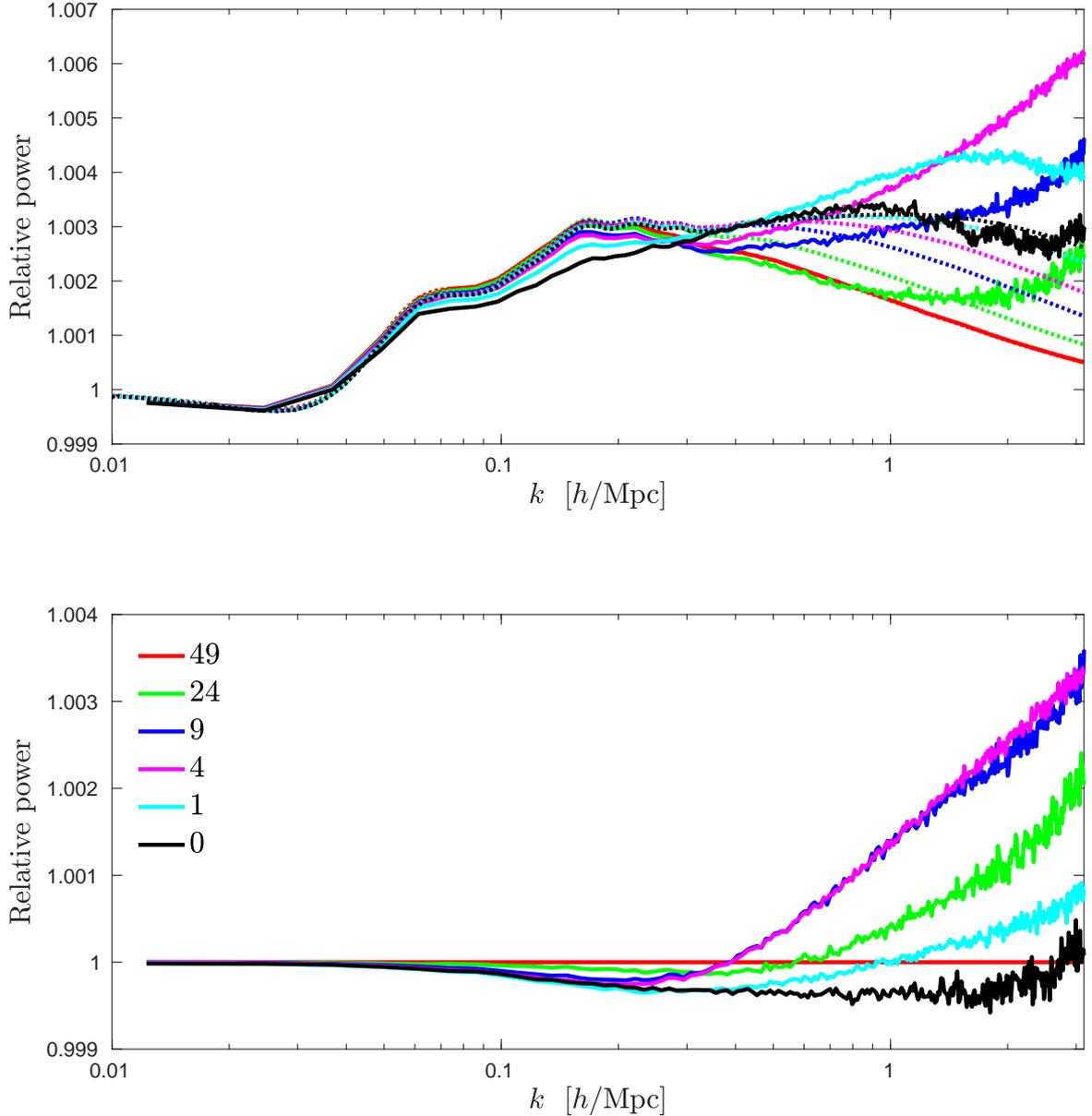}
\end{center}
 \vspace*{-4cm}
\caption{The top panel shows the relative power spectrum between a model with bosonic HDM and one with fermionic HDM for different redshifts. Linear theory is represented with dotted lines and non-linear theory with solid lines. The lower panel shows the relative power spectrum between models with BE and FD distributions in the $N$-body simulation, but where both $N$-body models have been initialised with the same fermionic TF.  All simulations have a HDM particle mass of $2{\rm eV}$.}
   \label{fig:nl_boson_2ev}
\end{figure}
\subsection{Fermions versus bosons}\label{sec:results_fvb}
As discussed in section \ref{sec:fermionvsboson} it is possible to approximately map a fermionic HDM particle into a bosonic particle and vice versa in linear theory. We have assessed the validity of this approximation in non-linear theory.

In Fig.~\ref{fig:velocity_dist2} we show the FD and BE cumulative probability distributions at $z=49$ (our $N$-body starting redshift) for a $2{\rm eV}$ HDM particle. The fermionic and bosonic temperatures are related by $T_{{\rm eff},f} = 6 T_{{\rm eff},b}/7$. With logarithmic axis, see the right hand panel, it can clearly be seen that the BE distribution has a much more pronounced tail at the low velocity end.

Fig.~\ref{fig:nl_boson_2ev} shows the linear and non-linear redshift evolution of the relative power spectrum between bosonic and fermionic models. Focusing on the upper panel, we see that even with our mapping procedure, the bosonic model initially gives a power spectrum which is larger by $0.3\%$. The redshift evolutions for the linear and non-linear calculations are markedly different, but they never become larger than $0.5\%$. At $z=0$ the linear and non-linear relative power spectra are virtually identical, differing by less that $0.1\%$ over all simulated scales.

The difference between linear and non-linear theory can be understood from the lower panel of Fig.~\ref{fig:nl_boson_2ev}. Here is shown the difference between two simulations which are both initialised with a TF from a fermionic cosmology, but in the $N$-body simulation, the HDM component is given either a BE or an FD thermal velocity component. Due to the larger low velocity tail of the BE distribution, the relative power spectrum initially increases at small scales. But as the gravitational potential deepens and the thermal velocity distributions redshift, a larger fraction of the particles in the BE and FD distributions can cluster at a given scale. Due to the crossover of these two distributions, see Fig.~\ref{fig:velocity_dist2}, the simulation with FD thermal velocities will eventually give rise to more structure, i.e.~the relative power spectrum falls below unity. It is worth noticing that this latter effect happens at large scales, $k\sim 0.2 h/{\rm Mpc}$, at high redshift.

Returning to the upper panel of Fig.~\ref{fig:nl_boson_2ev}, this pattern can roughly be identified by dividing the non-linear relative power spectrum with the linear relative power spectrum.

 \begin{figure}[t]
  \vspace*{-4.0cm}
\begin{center}
\hspace*{-1.0cm}
\includegraphics[width=18cm]{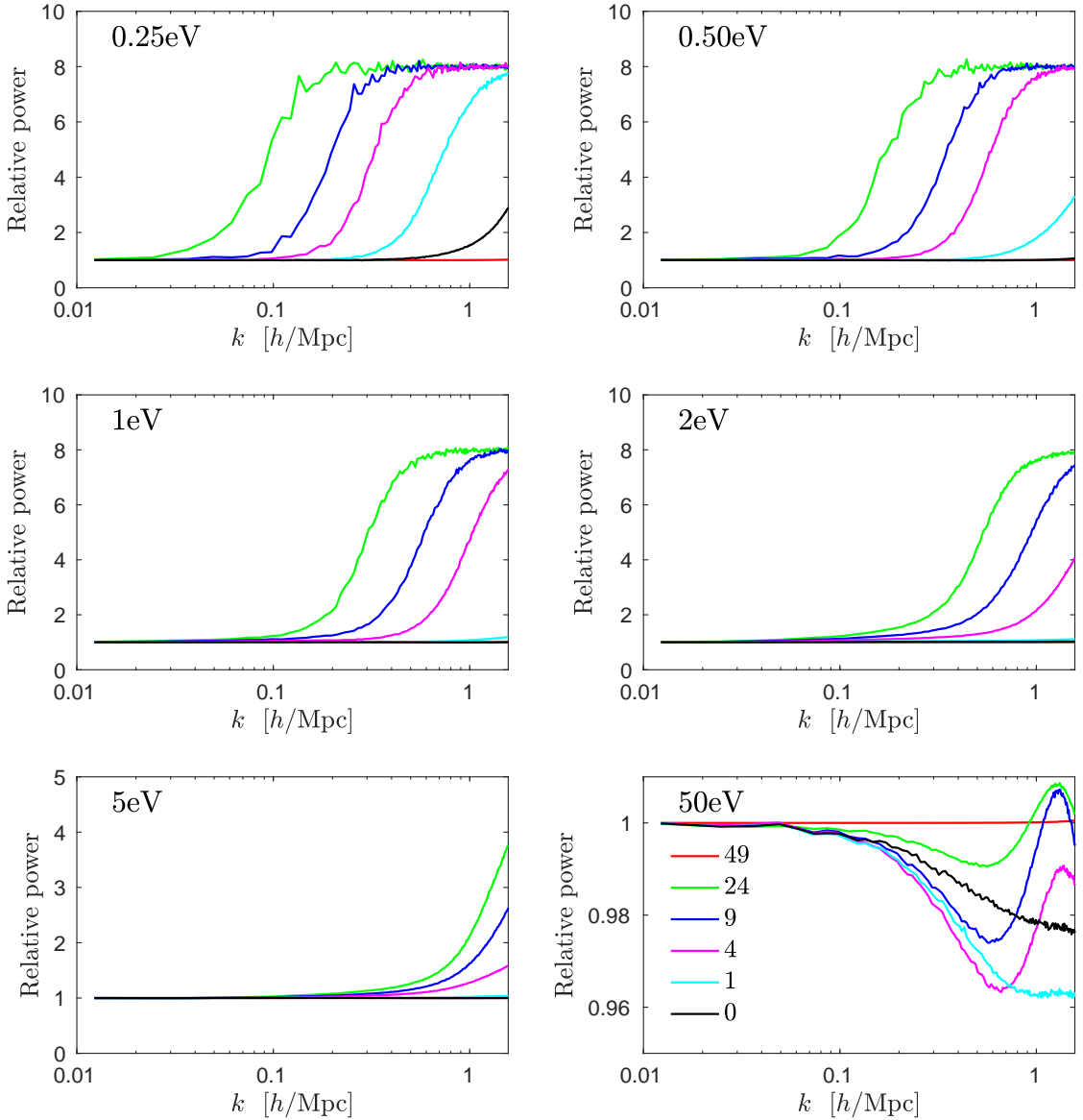}
\end{center}
 \vspace*{-4cm}
\caption{The panels show $P_{X}(512^3) / P_{X}(1024^3)$ for different redshifts, i.e.~the effect of changing the number of $N$-body particles. Note that the underlying gravitational potential differs by some percent due to the different number of CDM particles. This effect cannot be seen for the first 5 masses, where the thermal velocity is dominant, but it is clearly visible for the highest HDM mass.}
   \label{fig:nu_power_conv}
\end{figure}
\subsection{Convergence}
Fig.~\ref{fig:nu_power_conv} shows the ratio between the HDM power spectra for the $512^3$ and $1024^3$ particle simulations. Focusing on the $0.25 {\rm eV}$ case, the power spectra initially agree at $z=49$ but at $z=24$ they are vastly different with a constant offset of a factor of 8 (the relative number of HDM $N$-body particles) for $k\gtrsim 0.1 h /{\rm Mpc}$. At lower redshift, as the HDM $N$-body particle velocities redshift, this pattern is shifted towards larger wavenumbers. Physically, the HDM $N$-body particles are able to fall into the gravitational potentials, thereby reducing the white noise term in the power spectrum.

Increasing the HDM particle mass from $0.25{\rm eV}$ to $5{\rm eV}$ at a fixed redshift, the noise pattern is shifted towards smaller scales. But for the $50{\rm eV}$ case, the situation is markedly different. This can be understood from the following considerations: The $50{\rm eV}$ HDM $N$-body particles are basically cold at all simulated scales at our $N$-body starting redshift of 49. The relative power spectrum is therefore dominated by the evolution of the gravitational potential in the two simulations, and not significantly related to the thermal velocities. Since the two simulations use a different number of CDM $N$-body particles, and a different size of the particle-mesh grid on which the long-range gravitational force is calculated, the relative HDM power spectrum is therefore very similar to the corresponding relative CDM power spectrum. 

\section{Discussion and conclusions}\label{sec:conclusions}
In this paper we have quantified the effect of generic hot dark matter models on the cosmological matter power spectrum. We have used a thermalised fermion species as the dark matter particle, but with an effective temperature different from standard model neutrinos.
Additionally, we have verified that a bosonic hot dark matter species can be mapped very accurately to a fermionic species so that our approach is well suited to study any thermalised hot dark matter species.

We have found that for particle masses in excess of a few eV the non-linear evolution of hot dark matter is markedly different than that of standard model neutrinos. While standard model neutrino HDM models always exhibit an excess suppression of power in the non-linear regime (relative to a $\Lambda$CDM only simulation) as compared to the linear regime, the reverse is true for large physical HDM masses ($\gtrsim 1$eV).
Indeed, as the physical particle mass increases to 50eV the model becomes almost indistinguishable from $\Lambda$CDM with no neutrino mass at low redshift even though linear theory indicates a significant difference in fluctuation power. This behaviour can be understood to arise from the transfer of power from larger to smaller scales, in two models with similar large scale power but differing small scale power.

\begin{figure}[t]
  \vspace*{-1.0cm}
\begin{center}
\hspace*{-0.2cm}
\includegraphics[width=16cm]{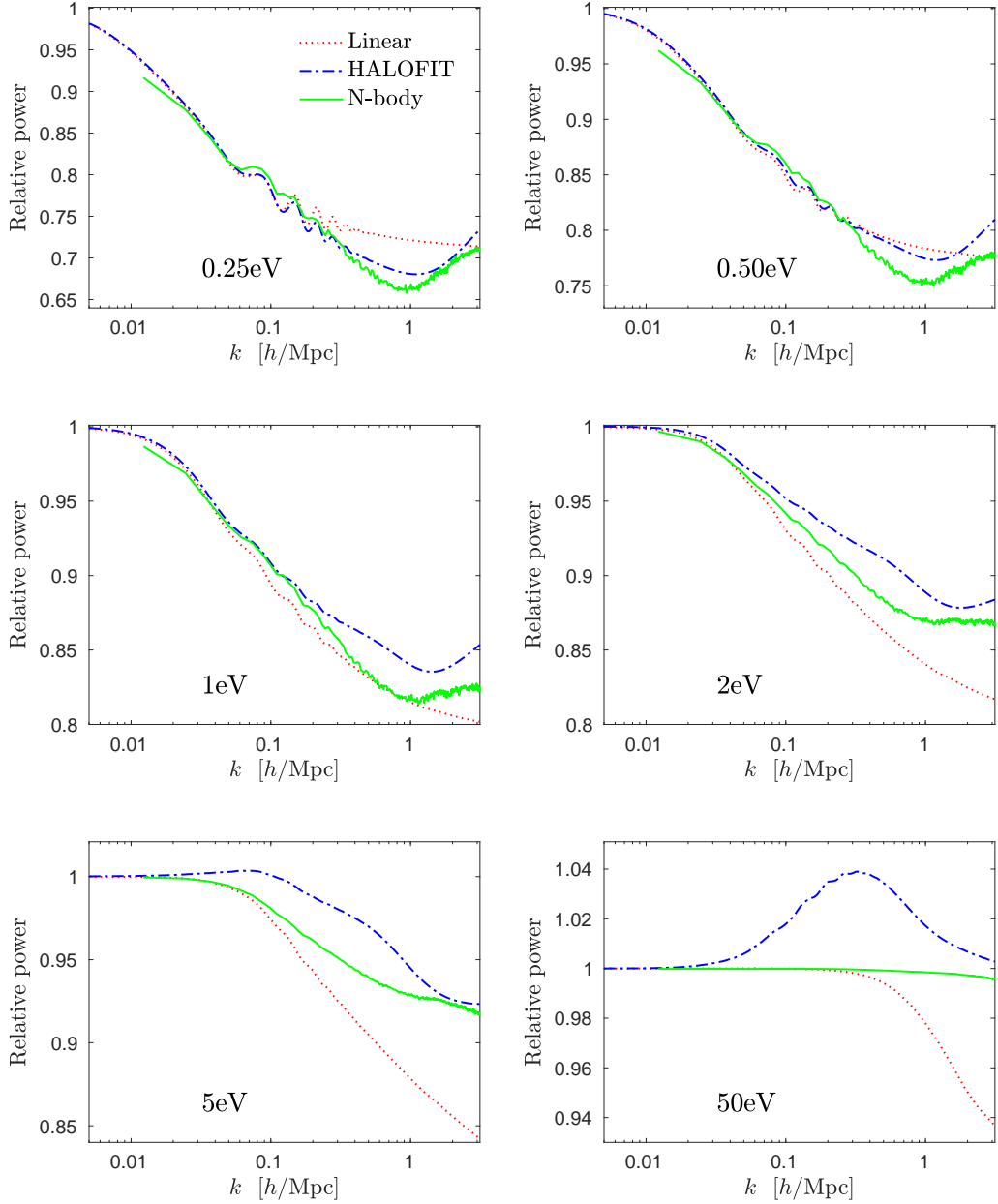}
\end{center}
 \vspace*{-2.5cm}
\caption{The figure shows the power spectrum suppression at $z=0$ between models with hot dark matter and one without. The suppression is calculated for 6 hot dark matter masses with both linear theory, HALOFIT and $N$-body simulations.}
   \label{fig:halofit}
\end{figure}

This behaviour of hot dark matter will be important to model properly when analysing future high precision cosmological data from surveys such as EUCLID \cite{Laureijs:2011gra}, where power spectrum observables can typically be measured at the 1\% level of precision for scales around $k \sim 1 \, h$/Mpc.
It also has implications for semi-analytic models of structure formation such as e.g.\ HALOFIT \cite{Smith:2002dz}. While HALOFIT has been 
recalibrated a number of times to include e.g.\ the effect of massive standard model neutrinos (see e.g.\ \cite{Bird:2011rb}), HALOFIT has not been tested for more general hot dark matter models where both the mass and the temperature of the hot dark matter component changes.
In Fig.~\ref{fig:halofit} we show the predictions of HALOFIT compared with linear theory and with the full $N$-body simulations for a variety of different particle masses. Clearly, for the smallest mass shown (0.25eV) the hot dark matter is sufficiently similar to standard model neutrinos that their effect on large scale structure is mapped fairly accurately by HALOFIT. However, for the larger particle masses HALOFIT clearly fails to properly model the effect of hot dark matter, and should be recalibrated using new $N$-body simulations.

\section*{Acknowledgements}
We thank T.~Tram for comments on the manuscript and acknowledge computing resources from the Danish Center for Scientific Computing (DCSC). This work was supported by the Villum Foundation.


\bibliographystyle{utcaps}

\providecommand{\href}[2]{#2}\begingroup\raggedright\endgroup

\end{document}